\documentclass[preprint,showpacs,amsmath,amssymb]{revtex4}
\usepackage{pstricks,graphicx,amsmath,bbm,mathrsfs,amssymb,psfrag,pifont}
\usepackage{color}
\def\hn{{\hat{n}}}
\def\hm{{\hat{m}}}
\def\ha{{\hat{a}}}

\begin{document}
\title{Measuring high-order photon-number correlations in multimode pulsed quantum states}
\author{Alessia Allevi}\email{alessia.allevi@uninsubria.it}
\affiliation{Dipartimento di Fisica e Matematica, Universit\`a degli Studi dell'Insubria and C.N.I.S.M., U.d.R. Como, I-22100 Como, Italy}
\author{Stefano Olivares}
\affiliation{Dipartimento di Fisica, Universit\`a degli Studi di Trieste, I-34151 Grignano (Trieste), Italy}
\affiliation{CNISM UdR Milano Statale, I-20133 Milano, Italy}
\author{Maria Bondani}
\affiliation{Istituto di Fotonica e Nanotecnologie, C.N.R., and C.N.I.S.M. U.d.R. Como,  I-22100 Como, Italy}
\date{\today}
\begin{abstract}
We implement a direct detection scheme based on hybrid photodetectors to experimentally investigate high-order correlations for detected photons by means of quantities that can be experimentally accessed. We show their usefulness in fully characterizing a multimode twin-beam state in comparison with classical states and, in particular, we introduce a nonclassicality criterion based on a simple linear combination of high-order correlation functions. Our scheme is self-consistent, allowing the estimation of all the involved parameters (quantum efficiency, number of modes and average energy) directly from the same experimental data. Results are in very good agreement with theory, thus suggesting the exploitation of our scheme for reliable state characterization in quantum technology.
\end{abstract}
\pacs{42.50.Dv, 42.50.Ar, 85.60.Gz}
\maketitle

Correlations play a fundamental role in the investigation of optical coherence \cite{mandel95}. Since their introduction \cite{glauber63}, correlation functions have been extensively investigated in connection to quantum state characterization, to define nonclassicality criteria \cite{vogel08}, and for the enhancement of ghost-imaging protocols \cite{chan09,chen10}. From the experimental point of view, the study of this topic dates back to the  pioneering work in which Hanbury Brown and Twiss discovered photon bunching in light emitted by a chaotic source \cite{brown56}. In the last decade, many experiments have been developed in which the photon-number correlations have been used to characterize the entangled states generated by parametric down-conversion \cite{rarity87,uren05,bussieres08,boiko09}. In all the mentioned cases, the detection was performed by means of single or arrays of avalanche photodiodes so that the possibility to recover the correlation of the number of photons was quite straightforward \cite{ivanova06,avenhaus10}, although the intensity range actually investigated by these systems was limited to much less than one mean photon \cite{kalashnikov11}. However, the investigation of more intense light beams is of extreme interest, especially in view of possible applications to quantum technology: pulsed optical states endowed with sizeable numbers of photons represent a useful resource as they are robust with respect to losses and their reliable experimental detection and characterization is relevant, especially for establishing nonclassicality.
\par
In this Letter we report on a direct detection scheme aimed at measuring high-order correlations by means of a pair of hybrid photodetectors, which are photon-counting detectors endowed with a good linear response up to tens of photons \cite{bondani09a}. In particular, we show that correlated bipartite optical states, namely a multimode twin-beam state (TWB) and a bipartite pseudothermal state generated in the mesoscopic photon-number domain \cite{allevi10a}, can be effectively detected and characterized. To this aim, we define and derive the analytical expression of correlation functions at any order by only using quantities that can be experimentally accessed by direct detection, taking into account the non-unit quantum efficiency of the detection scheme. We also show that high-order correlations represent a useful discriminating tool of the nature of the state and demonstrate that, at increasing correlation order, the differences between classical and quantum states become more and more evident.
\par
The correlation functions $g_\hn^{jk}$ are usually defined in terms of the normally-ordered creation and annihilation operators \cite{glauber65}:
\begin{equation}
 g_\hn^{jk}=\frac{\langle :\hn_1^j \hn_2^k:\rangle}{\langle :\hn_1:\rangle^j\langle :\hn_2:\rangle^k}=\frac{\langle \ha_1^{\dag j} \ha_1^{j} \ha_2^{\dag k} \ha_2^{k}\rangle}{\langle \ha_1^{\dag} \ha_1 \rangle^j\langle \ha_2^{\dag} \ha_2 \rangle^k}\ ,\label{eq:corrHIGHnormal}
\end{equation}
where $\ha_k$ is the field operator of the mode $k$-th and $\hn_k = \ha_k^{\dag} \ha_k$. $g_\hn^{jk}$ have a well-recognized meaning in connection with coherence properties of light and $n$-photon absorption process \cite{klyshko11}.
However, in a realistic direct detection scheme, we have only access to the shot-by-shot detected photons. We thus need to build analogous correlation functions for directly measurable quantities, namely:
\begin{equation}
 g_\hm^{jk}=
\frac{\langle \hm_1^j \hm_2^k\rangle}
{\langle \hm_1\rangle^j\langle \hm_2\rangle^k}\ ,\label{eq:corrHIGH}
\end{equation}
where $\hm_{k}$ is the operator describing the actual number of detected photons in the $k$-th arm of the bipartite state. Indeed, the functions $g_m^{jk}$ can be linked to the $g_\hn^{jk}$ of Eq.~(\ref{eq:corrHIGHnormal}), provided that a suitable description of the actual operation performed by the detector is given. If we assimilate the real detection to a Bernoullian process having efficiency $\eta$, we can express all the operatorial moments of the detected-photon distribution as a function of those of the photon distribution, i.e., $\hm_k^p = \sum_{h=1}^p c_h(\eta) \hn_k^h$, where the coefficients $c_h(\eta)$ are given in Ref.~\cite{agliati05}. Thus, by inserting this expression in Eq.~(\ref{eq:corrHIGH}) and exploiting commutation rules, we can express detected-photon correlations in terms of normally-ordered ones, i.e.,  $g_\hm^{jk}=\sum_{s,t=0}^{j,k} \varepsilon_{s,t}\, g_\hn^{st}$, where $g_\hn^{00}\equiv 1$ and $\varepsilon_{s,t}$ depend on the physical parameters of the system under investigation.
\par
On the other hand, the expected results for correlations should be evaluated by taking into account all the realistic experimental conditions, that is not only imperfect detection and imbalance of the arms of the bipartite state, but also its multimode nature. Thus, we need to extend the theoretical description of the detection process to the case of multimode states, for which all the $\mu$ modes in the field are measured shot-by-shot.
\par
We consider a multimode TWB $\left|\psi_{\mu}\right.\rangle = \otimes_{k=1}^{\mu} \left|\psi\right.\rangle_k$ in which each of the $\mu$ modes is in the same state, \emph{i.e.},
$\left|\psi\right.\rangle_k = \sum_{n} \langle \hn \rangle^{n}/(1+\langle \hn \rangle)^{n+1} \left|n\right.\rangle_k \otimes \left|n\right.\rangle_k$, $\forall k$, and, thus, is equally populated with the same average number of photons per mode $\langle \hn \rangle = \langle \hn_k \rangle$, $\forall k$ \cite{mandel95}. By exploiting the pairwise correlations, we can state that the overall number of photons in each shot is the same in the two arms, being the sum of $\mu$ equal contributions coming from the $\mu$ modes. We can thus write the multimode TWB in the following compact form
$\left|\psi_{\mu}\right.\rangle= \sum_{n=0}^\infty \sqrt{p^\mu_n}
\left|n^{\otimes}\right.\rangle\otimes\left|n^{\otimes}\right.\rangle$ ,
where $\left|n^{\otimes}\right.\rangle = \delta(n-\sum_{h=1}^{\mu}n_h)\,\otimes_{k=1}^{\mu}\left|n_k\right.\rangle_k$ represents the overall $n$ photons coming from the $\mu$ modes that impinge on the detector and:
\begin{equation}
p^\mu_n= \frac{\left(n +\mu-1\right)!}{{n!\left(\mu - 1 \right)! \left(\langle \hn\rangle/\mu+1 \right)^{\mu} \left(\mu/\langle \hn\rangle+1 \right)^{n}}}
\end{equation}
is the photon-number probability distribution for the multimode TWB. As one may expect, we obtain a final result that depends only on the number of modes $\mu$, the mean value of the number of photons $\langle \hn\rangle$, and the overall detection efficiencies of the two detection chains, $\eta_1$ and $\eta_2$. In the case of $\eta_1 = \eta_2$ and considering correlation up to  the 4-th order, we have (we put $g^{hk} \equiv g^{hk}_\hm$):
\begin{subequations}\label{eq:teoTWB}
\begin{align}
 g^{11} &= G^{1}_\mu+\frac{\eta}{\langle \hm\rangle}\\
 g^{21} &= g^{12} = G^{2}_\mu+
 G^{1}_\mu\frac{1+2\eta}{\langle \hm\rangle}+
 \frac{\eta}{\langle \hm\rangle^2} \\
 g^{22} &= G^{3}_\mu+
 2G^{2}_\mu\frac{1+2\eta}{\langle \hm\rangle}
 +G^{1}_\mu\frac{1+4\eta+2\eta^2}{\langle \hm\rangle^2}+
 \frac{\eta}{\langle \hm\rangle^3}\\
 g^{31} &= g^{13} =G^{3}_\mu+
 3G^{2}_\mu\frac{1+\eta}{\langle \hm\rangle}
 + G^{1}_\mu\frac{1+6\eta}{\langle \hm\rangle^2}+
 \frac{\eta}{\langle \hm\rangle^3},
\end{align}
\end{subequations}
where $\langle \hm\rangle$ is the average number of detected photons and {$G^{k}_\mu = \prod_{j=1}^{k}(j+\mu)/\mu$. Indeed, the previous equations also describes two other classes of states: by choosing $\eta=0$ we obtain the correlation functions for a multimode thermal state, while for $\eta=0$ and $\mu \rightarrow \infty$ we get the results for the coherent state \cite{brida11}.
\par
\begin{figure}[h]
\includegraphics[width=8cm]{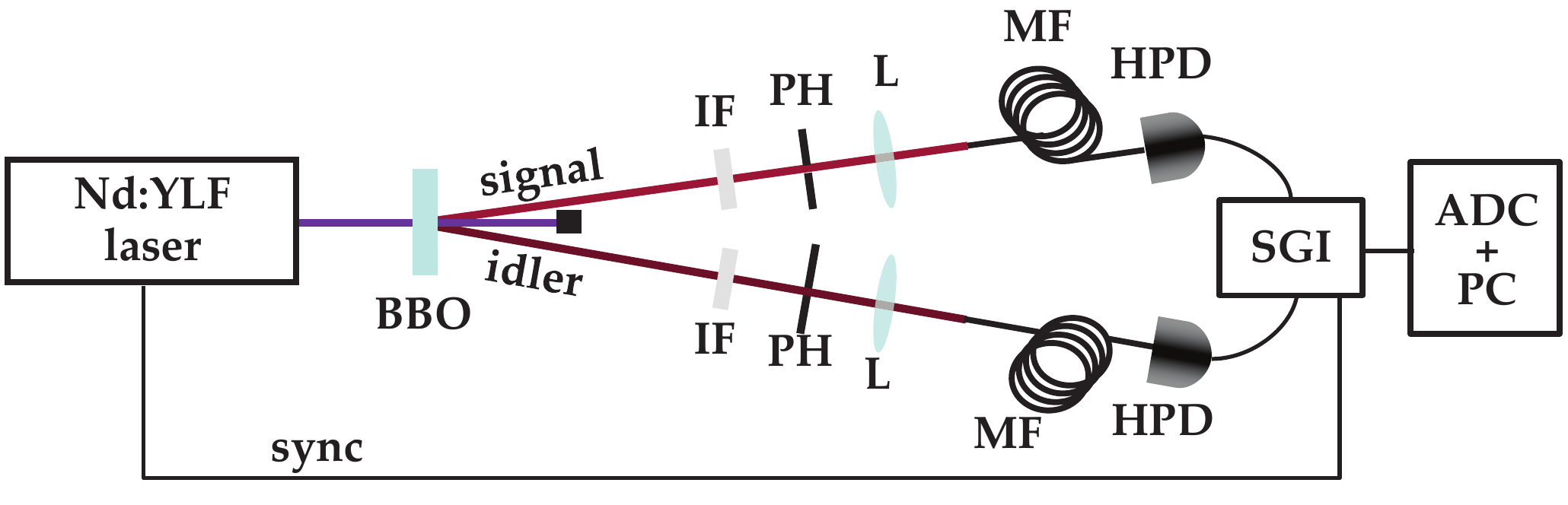} 
\caption{Scheme of the experimental setup. See the text for details.}\label{f:setup}
\end{figure}
We generated a multimode TWB by the third harmonics (349~nm wavelength) of a mode-locked Nd:YLF laser regeneratively amplified at 500~Hz (High-Q Laser Production, Austria) impinging on a type-I BBO crystal ($\beta$-BaB$_2$O$_4$, Castech, China, cut angle 34 deg, 4-mm thick). According to the experimental setup sketched in Fig.~\ref{f:setup}, we adopted a non-collinear interaction geometry in order to avoid possible residues of the pump beam. To obtain a good balancing of the quantum efficiencies, we selected two portions of the signal and idler cones close to frequency degeneracy, namely at 690~nm and 706~nm, by using two interference filters (IF, in Fig.~\ref{f:setup}). The collection of a single coherence area in the two parties of the TWB state was obtained by inserting two pin-holes (PH, 2~mm diameter) at 107~cm and 109.5~cm from the BBO, respectively. The light passing the pin-holes was delivered through two multimode optical fibers to two hybrid photodetectors (HPD, R10467U-40, Hamamatsu, Japan), which are detectors endowed with a partial photon-counting capability and a good linear response up to 100 photons. Their outputs were amplified (preamplifier A250 plus amplifier A275, Amptek), synchronously integrated (SGI, SR250, Stanford) and digitized (AT-MIO-16E-1, National Instruments). Each experimental run was performed on 50~000 subsequent laser shots at different values of the pump intensity. The self-consistent procedure to analyze the outputs of each detection chain is explained in details in Refs.~\cite{bondani09a,andreoni09}, here we only remind that the procedure allows us to obtain the mean value of detected photons $\langle \hm \rangle$, the number of modes $\mu$, and the quantum efficiency $\eta$ directly from the experimental data.
\par
In previous works, we have already demonstrated that with the present detection apparatus assisted by the self-consistent analysis method it is possible to reconstruct both the detected-photon statistics \cite{bondani09b} and the shot-by-shot second-order correlation in the number of detected photons \cite{allevi10b}. Moreover, we have shown that the nonclassical nature of the TWB state can be proven by evaluating the noise reduction factor \cite{allevi10a}. Nevertheless, as our TWB state is intrinsically multimode ($\mu\approx 100$) and the effective detection efficiency is rather low ($\eta\approx 5~\%$), the experimental characterization of the state in terms of sub-shot-noise correlations becomes challenging. On the contrary, the calculation of high-order correlations offers the possibility to better discriminate the nature of the state under investigation also in critical situations.
\begin{figure}[h]
\includegraphics[width=0.33\textwidth]{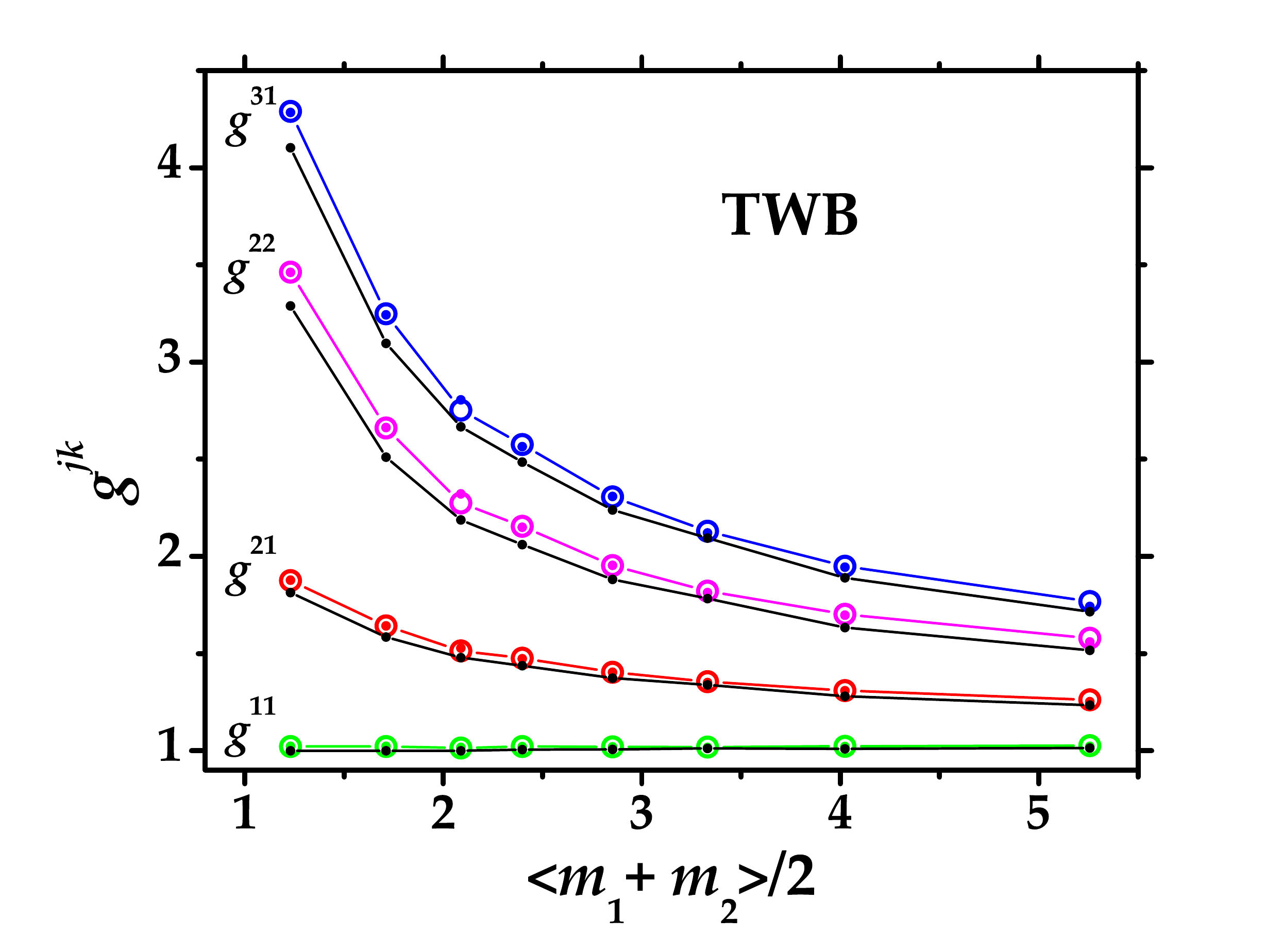} 
\vspace{-0.3cm}
\caption{High-order correlation functions for a multimode TWB state as a function of $\langle \hm_1+\hm_2 \rangle/2$. Full circles: experimental data; Line+open symbols: theoretical expectations; Line+symbol: theoretical prediction for a multimode thermal state.} \label{f:experCORR}
\end{figure}
In Fig.~\ref{f:experCORR} we plot the experimental data (full circles) obtained by evaluating $g^{jk}$ up to the fourth order together with the theoretical expectations (line+open circles) for a multimode TWB calculated according to Eqs.~(\ref{eq:teoTWB}) by using the values of $\langle \hm \rangle$, $\mu$ and $\eta$ as directly obtained from the experimental data \cite{agliati05,allevi10a}. Since the values $\langle \hm_1 \rangle$ and $\langle \hm_2 \rangle$ are slightly different, in Fig.~\ref{f:experCORR} and in all the other plots we use the quantity $\langle \hm_1 + \hm_2 \rangle/2$. The agreement of experimental data with theory is very high. For comparison, in Fig.~\ref{f:experCORR} we also plot the theoretical predictions for a classically-correlated multimode thermal state having the same mean value and number of modes (line+symbols) \cite{brida11}.
It is apparent that quantum correlations are always higher than classical ones, even if the absolute value of the difference is rather low due to the large number of modes and to the low value of quantum efficiency. Moreover, as the order of correlations increases, the difference becomes larger.
\par
\begin{figure}[h]
  \includegraphics[width=0.33\textwidth]{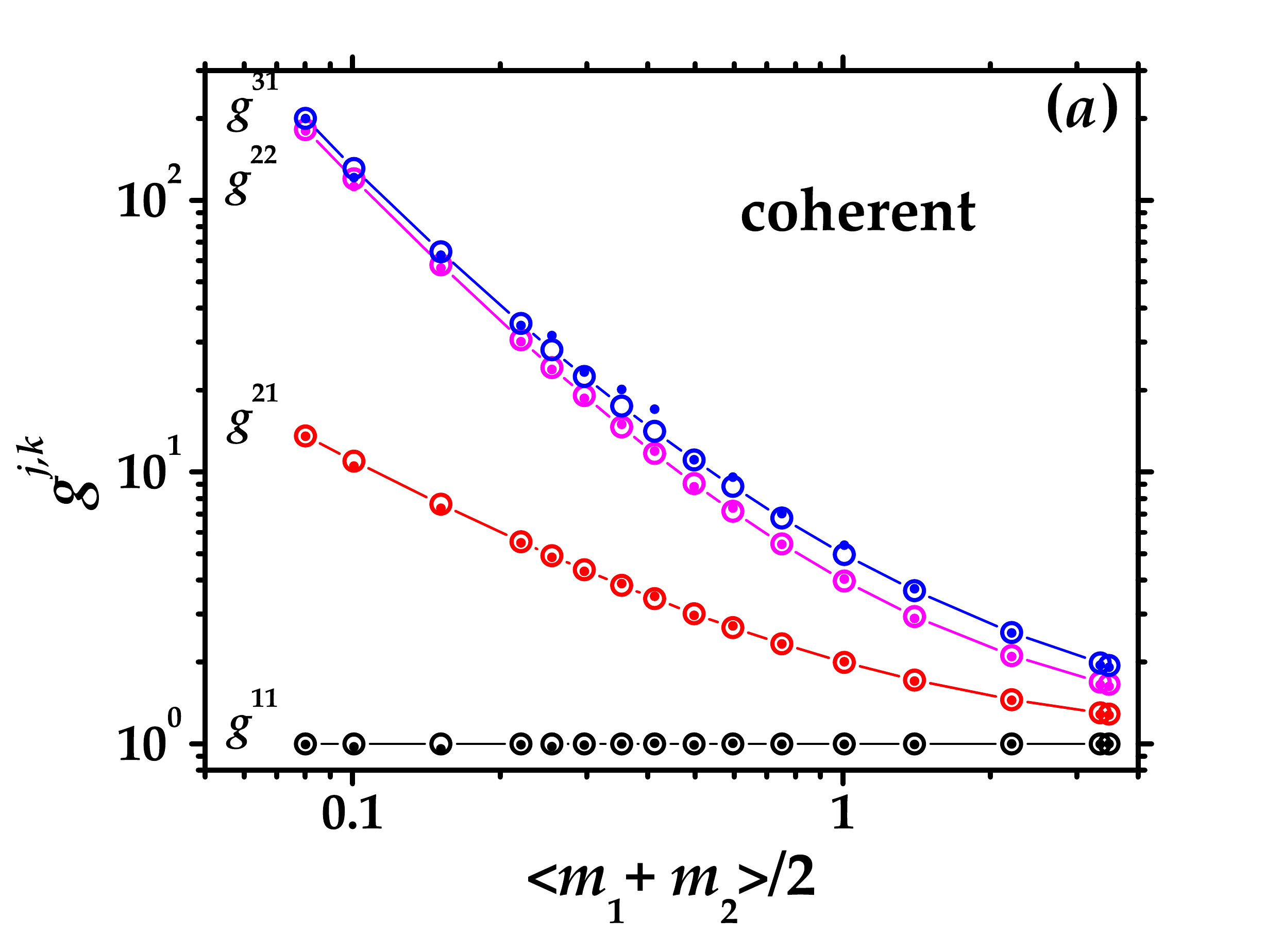}
  \vspace{-0.3cm}

  \includegraphics[width=0.33\textwidth]{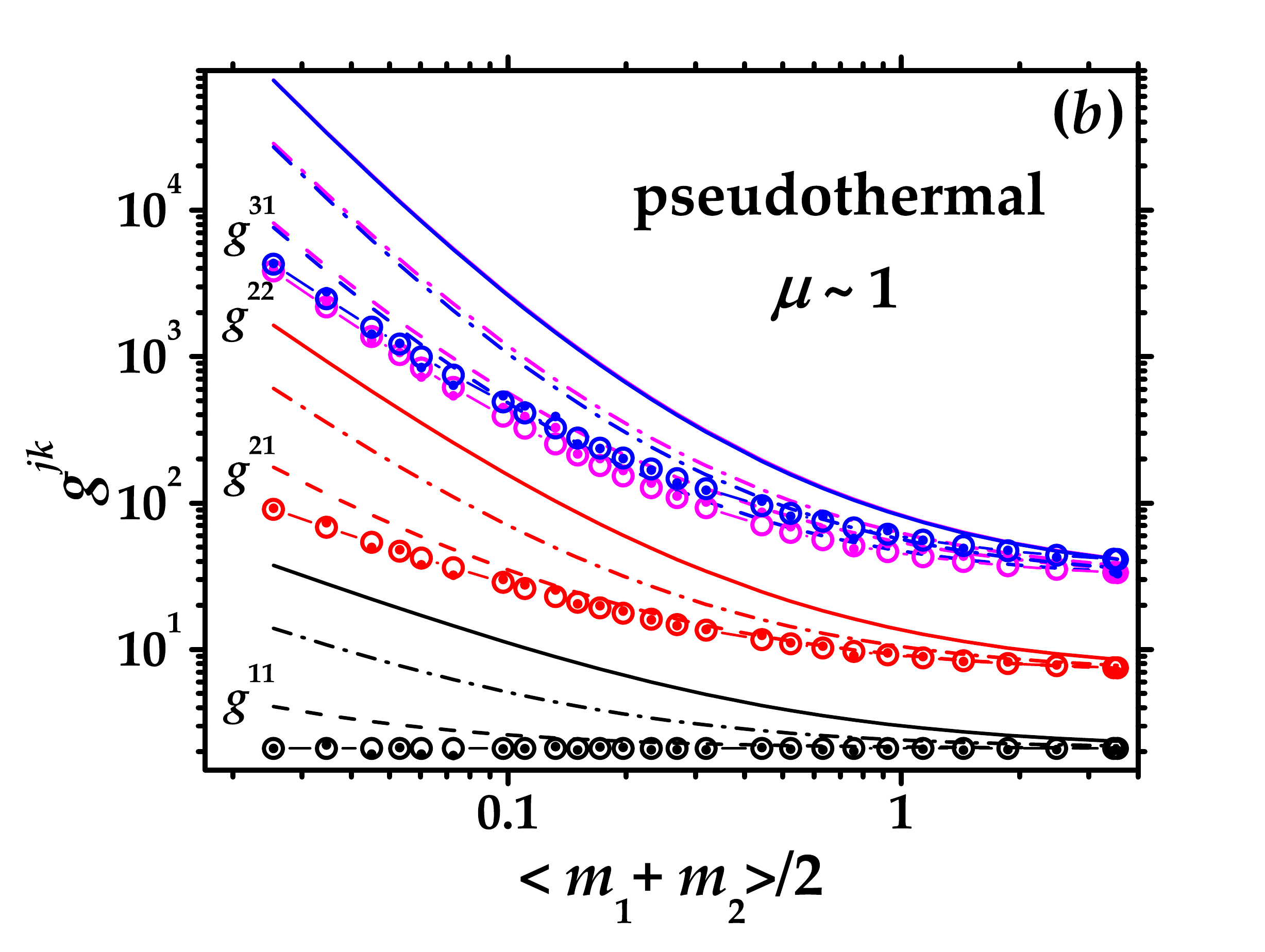}
\vspace{-0.3cm}

  \includegraphics[width=0.33\textwidth]{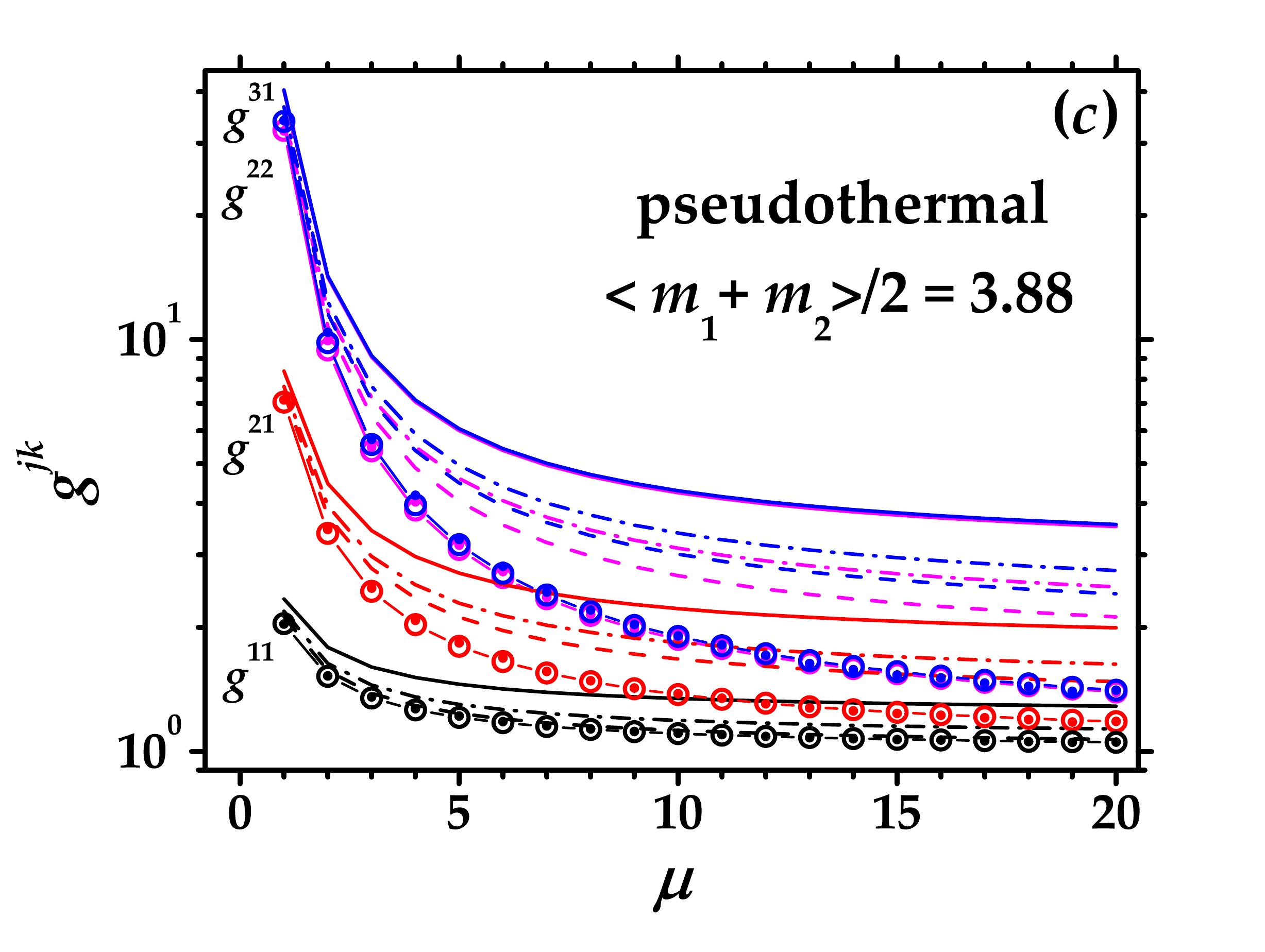}
  \vspace{-0.3cm}
  \caption{High-order correlation functions for coherent state ($a$) and single-mode pseudothermal state ($b$) as a function of $\langle \hm_1+\hm_2 \rangle/2$ and multimode pseudothermal state ($c$) as a function of the number of modes. Full circles: experimental data; Line+open symbols: theoretical expectations. In panels ($b$) and ($c$) theoretical predictions for a TWB state with same parameters measured with different quantum efficiencies ($\eta=5~\%$, dashed line, $\eta=30~\%$, dash-dotted line and $\eta=90~\%$, full line). Note the logarithmic scales in the plots.}   
  \label{f:thermal}
\end{figure}
In order to explore the capability of our system in different regimes, we also measured a coherent state and single and multimode pseudothermal states generated by a rotating ground-glass plate and divided at a beam splitter \cite{allevi10b,allevi10c}. In Fig.~\ref{f:thermal} we plot the measured high-order correlations for some of these states. In panels ($a$) and ($b$), we plot the experimental results (full circles) and the theoretical expectations (line+open symbols) for the coherent state and for a single-mode thermal state as a function of the mean value of the number of detected photons on the two components of the bipartite state. In panel ($c$) we plot the experimental results (full circles) and the theoretical expectations (line+open circles) for a multimode thermal state at fixed mean value $\langle \hm_1+\hm_2 \rangle/2 =3.88$ as a function of the number of modes $\mu$. All the experimental data superimpose to the theoretical expectations. As a comparison, in panels ($b$) and ($c$) we plot the value expected for a TWB state having the same physical parameters and measured with different quantum efficiencies.
We note that the expected correlations for TWB are always higher than for classically correlated pseudothermal states, and the difference increases at increasing quantum efficiency. At fixed quantum efficiency, the maximum difference is achieved at any order for low values of both $\langle \hm \rangle$ and $\mu$.
We can conclude that by measuring high-order correlations we can achieve a better discrimination between quantum and classical correlations even at a very low quantum efficiency.
\par
A more direct way to establish nonclassicality is to address a suitable parameters satisfying a boundary condition \cite{klyshko96,short83,vogel06}. In our experiment, as a first check, we verify the Schwarz inequality:
\begin{equation}
\langle \hm_1 \hm_2\rangle/\sqrt{\langle \hm_1^2\rangle \langle \hm_2^2\rangle}>1\,,\label{eq:schw}
\end{equation}
the noise reduction factor:
\begin{equation}
\left[\langle (\hm_1-\hm_2)^2\rangle  - \langle \hm_1-\hm_2\rangle^2\right]/\langle \hm_1+\hm_2\rangle<1\,,\label{eq:subshot}
\end{equation}
and the new inequality:
\begin{equation}
\langle \hm_1\rangle \langle \hm_2\rangle\,
\frac{g^{22} - [g^{13}]_{\rm s}}{g^{11}}+
\sqrt{\langle \hm_1\rangle \langle \hm_2\rangle}\,\, \frac{[g^{12}]_{\rm s}}{g^{11}}>1\,,\label{eq:ineq}
\end{equation}
in which we have introduced symmetrized quantities $[g^{hk}]_{\rm s} = \frac12 (g^{hk}+g^{kh})$ to take into account all the unavoidable asymmetries of the physical system. Indeed, all the above inequalities must be fulfilled by nonclassical light. Note that Ineq.~(\ref{eq:ineq}) is equivalent to require $g_\hn^{22}-g_\hn^{31}>0$ in the case of perfectly balanced efficiencies.
\begin{figure}[h]
\includegraphics[width=0.33\textwidth]{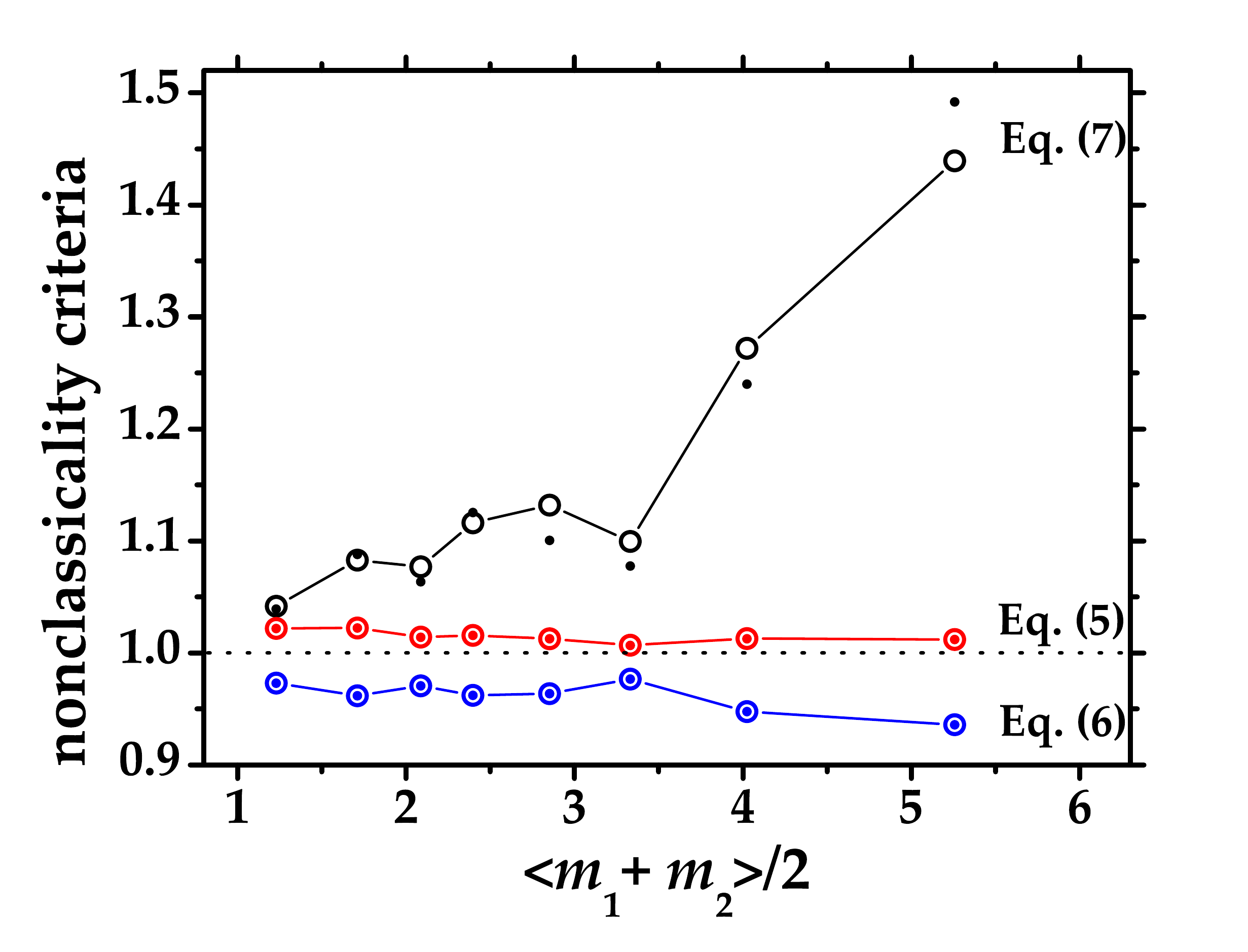}
\vspace{-0.3cm}
\caption{Nonclassicality criteria for a multimode TWB as a function of the mean value of the number of detected photons. Full circles: experimental data; Line+open symbol: theoretical prediction.} \label{f:ineq}
\end{figure}
In Fig.~\ref{f:ineq} we plot the results for the three inequalities of above for our experimental data on multimode TWB. We note that, as expected, the measured state is nonclassical according to all the three inequalities, but the amount of nonclassicality, that we can quantify with respect to the boundary unitary value, is much larger for the condition in Ineq.~(\ref{eq:ineq}). This indicates that using high-order correlations can be helpful in all the situations in which the nonclassical nature is critical to be proven by other criteria, such as the case of a TWB state endowed with a high number of modes and/or a low quantum efficiency, for which we expect very small values of noise reduction factor.
\par
In conclusion, we have defined correlation functions at any order by means of quantities that can be experimentally accessed by direct detection with an experimental apparatus that allows accessing sizeable mean photon values. We have introduced a nonclassicality criterion based on these high-order correlations that represents a useful discriminating tool of the nature of the state in critical cases, in which other criteria are violated by a small amount: in some sense this criterion acts as an ``amplifier'' of nonclassicality violation. We have tested our theoretical and experimental procedure on a multimode TWB state by developing a multimode description that makes the calculation of high-order correlations straightforward. For the sake of completeness, we have also presented the direct comparison with an equally populated multimode pseudo-thermal state. The experimental results are in very good agreement with the theory, thus encouraging the exploitation of our setup for the reliable experimental characterization of quantum states for quantum technology.
\par
The authors thank M.~G.~A.~Paris for fruitful discussions. SO acknowledges financial support from the University of Trieste through the ``FRA 2009''.

\vfill

\end{document}